\def\ptitle{Exact and approximate solutions to Schr\"odinger's equation with decatic potentials}

\documentclass{revtex4}
\usepackage{amsmath, latexsym}
\usepackage{graphicx}
\usepackage{rotating}%\usepackage{amsmath}
\usepackage{amssymb}

\newcommand{\schro}{Schr\"{o}dinger }

\newtheorem{theorem}{Theorem}
\newtheorem{definition}{Definition}
\begin{document}

% -----------------------------------------------
%\hspace*{3.2 in}CUQM-XXX
%\hspace*{4 in}[scoul] [9 Nov 2009]
%\vspace*{0.4 in}
% -----------------------------------------------

\title{\ptitle}

\author{David Brandon and Nasser Saad}

\affiliation{Department of Mathematics and Statistics,
University of Prince Edward Island, 550 University Avenue,
Charlottetown, PEI, Canada C1A 4P3.}

\email{dbrandon@upei.ca} \email{nsaad@upei.ca}
\medskip
\begin{abstract}
\noindent{\bf Abstract:}  The one-dimensional Schr\"odinger's equation  is analysed with regard to the existence of exact solutions for decatic polynomial potentials. Under certain conditions on the potential's parameters, we show that the  decatic polynomial potential $V(x)=ax^{10}+bx^8+cx^6+dx^4+ex^2$, $a>0$  is exactly solvable. By examining the polynomial solutions of certain linear differential equations with polynomial coefficients, the necessary and sufficient conditions for corresponding energy-dependent polynomial solutions are given in detail. It is also shown that these polynomials satisfy a four-term recurrence relation, whose real roots are the exact energy eigenvalues.  Further, it is  shown that these polynomials generate the eigenfunction solutions of the corresponding \schro equation. Further analysis for arbitrary values of the potential parameters using the asymptotic iteration method is also presented.
 \end{abstract}
 \maketitle

\noindent{\bf PACS :} 02.30.Hq, 03.65.Ge, 03.65.-w.
\vskip 0.1 true in

\noindent{\bf Keywords:} Differential equations, anharmonic oscillators, polynomial potentials, decatic potentials, energy spectrum, exact solutions.

\maketitle
% ----------------------------
\section{Introduction}
% ----------------------------
\noindent The problem of finding analytic (polynomial) solutions of linear differential equations has lost some of its interest in recent years. The lack of interest is mainly due to the vast development of numerical methods and the fact that much of their study has been superseded by more general work based on the theory of Lie algebra \cite{kal, kam,tur,tur1,fin,gon, Zn4}. Recently, the interest of analytic (polynomial) solutions has been renewed by the analysis of confined and un-confined
quantum systems and also by the search of closed-form solutions to (Schr\"odinger-type) differential equations with polynomial coefficients \cite{adam,azad,zhang1,zhang2,zhang3,cas,fra, jac, bender, carl,kr2,  bo, hi,hi1,de, dong,dong1,dong2,ym,asi,
saad1,saad8, saad2,saad3,saad4, ere,tur, voros,nap,Zn1,Zn2,Zn3, Zhang}. A reason for such interest is that in many problems in quantum mechanics, especially
those arising from Schr\"odinger's equation after the separation of the asymptotic-behaviour factor of the wave function,
there remains a polynomial-type factor in the solution \cite{azad, kam,tur,tur1,fin,gon,Al,roy,ko,kal,hi,fra,carl,ere,ym, saad0,adh,de,saad1,saad2,saad3,saad4,saad5,saad6,saad7,saad8,web, zhang1,
zhang2,zhang3, Zhang}. One can further  argue that analytic (polynomial) solutions provide a deeper quantitative insight into the physical model under investigation and in many cases makes the conceptual understanding of physics straightforward and sometimes intuitive \cite{Al,fra2,fle,mag,mam}. Moreover, these
solutions (if they are available) are  valuable tools for checking and improving numerical methods
introduced for solving complicated physical systems \cite{Al,hi1}.  
\vskip0.1true in
\noindent In the present work we study the exact and approximate solutions of the  Schr\"odinger equation with decatic polynomial potential \cite{ere, zhang1,zhang2, dong, feng} 
\begin{equation}\label{eq1}
	\hat{H}\psi=-\frac{d^2\psi}{dx^2}+V(x)\psi=E\psi,\quad V(x)= ax^{10}+bx^8+cx^6+dx^4+ex^2\qquad (a>0,~-\infty<x<\infty).
\end{equation}
The importance of this type of even-power potentials follows from its relevance to different models of charmonium system \cite{fle,mam} and its connection with the analysis of anharmonic and double-well oscillators \cite{cha,mee,fra2,dong2,mo,gu,fle,li,mag,feng,roy,mam,er}. We show under certain relations between the potential parameters $a$, $b$, $c$, $d$, and $e$, we have exact solutions for the \schro equation  \eqref{eq1}, in the sense that $\psi$ and $E$ are completely determined. These relations are expressed in terms of energy-dependent polynomials that satisfies a four-term recurrence relation. We show further that these polynomials generate the corresponding eigenfunctions. Our approach depends on analysing the analytic (polynomial-type) solutions of a second-order differential equation
\begin{equation}\label{eq2}
\left(\sum_{k=0}^6 a_{6,k} x^{6-k}\right)y''+\left(\sum_{k=0}^5a_{5,k}x^{5-k}\right)y'-\left(\sum_{k=0}^4 \tau_{4,k}x^{4-k}\right)y=0,\qquad(a_{6,0}^2+a_{5,0}^2\neq 0),
\end{equation}
where $a_{6,k},a_{5,k}$ and $\tau_{4,k}$, for all $k$, are constants independent of the variable $x$. Thereby, we give the necessary and sufficient conditions for its polynomial solutions $y=\sum_{k=0}^n c_k x^k$. In the second part of the present study, we introduce the asymptotic iteration method \cite{saad1} to obtain accurate approximate solutions for the decatic potential  \eqref{eq1} with arbitrary values of the potential parameters. In section II, we set up the Schr\"odinger equation for the decatic-power potential \eqref{eq1} and we show it reduces to a differential equation of a  type similar to equation \eqref{eq2}. In section III, we provides both the necessary and sufficient conditions for the existence of polynomial solutions of the (general) linear differential equation \eqref{eq2}. In section IV, we  examine the exact and approximate solutions of decatic potential \eqref{eq2} and report some numerical results illustrate the usefulness of our approach where we compare our results with the existing literature. 
%%%%%%%%%%%%%%%%%%%
\section{One-dimensional decatic-power potential}
%%%%%%%%%%%%%%%%%%%
\noindent 
In this section, we consider the one-dimensional Schr\"odinger equation 
\begin{equation}\label{eq3}
	-\psi''(x)+(V(x)-E)\psi(x)=0,\qquad V(x)=ax^{10}+bx^8+cx^6+dx^4+ex^2,~-\infty<x<\infty,
\end{equation}where $a,b,c,d$, and $e$ are real constants with $a>0$.
Assume the solution has the form
\begin{equation}\label{eq4}
	\psi(x)=\chi(x)\cdot e^{-\phi(x)} \qquad\Big(\lim\limits_{|x|\rightarrow \infty}\phi(x)=\infty\Big).
\end{equation}
Substituting this expression into equation \eqref{eq3} gives a second-order linear differential equation for $\chi(x)$ as
\begin{equation}\label{eq5}
	\chi''(x)=2\phi'(x)\chi'(x)+\Big(\phi''(x)-\big(\phi'(x)\big)^{2}+V(x)-E\Big)\chi(x).
\end{equation}
Without loss of generality, we may assume, for the nodeless eigenstate $\psi_0(x)$, that $\chi(x)=1$.  In which case, equation \eqref{eq5} reduces to Riccati's equation
\begin{equation}\label{eq6}
	\phi''(x)=E-V(x)+\big(\phi'(x)\big)^{2}\quad \text{or }\quad u'(x)=E-V(x)+u^2(x), \quad \text{where}\quad u(x)=\phi'(x).
\end{equation}
\begin{definition} (\cite{RAIN}, page 474) By the symbol $\left[\sqrt{P(x)}\right]$, where $P(x)$ is a polynomial of even-degree, we shall mean the polynomial part of the expansion of $\sqrt{P(x)}$ in a series of descending integral powers of $x$.
\end{definition}
\begin{theorem}  (\cite{RAIN}, Theorem 1, page 474) 	If in $
		{du}/{dx}=A_{0}+u^2$, 
	$A_0\equiv A_{0}(x)$ is a polynomial of even degree, then no polynomial other than
\begin{equation}\label{eq7}
		u=\pm\left[\sqrt{-A_{0}}\right]
\end{equation}
	can be a solution of \eqref{eq6}.  If the degree of $A_{0}$ is odd, there is no polynomial solution of \eqref{eq6}.
\end{theorem}
\noindent By means of this theorem, we obtain for the solution of equation \eqref{eq3} 
\begin{equation}\label{eq8}
	\psi(x)=\chi(x)\cdot \exp\left(-\frac{\sqrt{a}}{6}x^6-\frac{b}{8\sqrt{a}}x^4+\frac{\left(b^2-4ac\right)}{16a^{3/2}}x^2\right),
\end{equation}
where $\chi(x)$ satisfy the second-order linear differential equation
\begin{align}\label{eq9}
\chi''(x)&-\left(2\sqrt{a}x^5+\frac{b}{\sqrt{a}}x^3-\frac{\left(b^2-4ac\right)}{4a^{3/2}}x\right)\chi'(x)\notag\\
&-\left(\dfrac{4ac-b^2}{8a^{3/2}}-E+\dfrac{64a^3e+96a^{5/2}b-b^4+8ab^{2}c-16a^2c^2}{64a^3}x^2+
				\dfrac{8a^2d+40a^{5/2}+b^3-4abc}{8a^2}x^4\right)\chi(x)=0.
\end{align}

%%%%%%%%%%%%%%%%%%%%%%%%%%%%%%%%%%%%%%
\section{Necessary and sufficient conditions for polynomial solutions}
%%%%%%%%%%%%%%%%%%%%%%%%%%%%%%%%%%%%%%%%%%%%%%%%%%
\noindent In this section, we study a class of  differential equations that generalized equation \eqref{eq9} and  
we establish the necessary and sufficient conditions for the existence of polynomial solutions.

\begin{theorem}\label{conds}
The necessary condition for the existence of polynomial solutions of the differential equation  
\begin{align}\label{eq10}
(a_{6,0}x^6+a_{6,1}x^5&+a_{6,2}x^4+a_{6,3}x^3+a_{6,4}x^2+a_{6,5}x+a_{6,6})y''\notag\\
+(a_{5,0}x^5&+a_{5,1}x^4+a_{5,2}x^3+a_{5,3}x^2+a_{5,4}x+a_{5,5})y'-(\tau_{4,0}x^4+\tau_{4,1}x^3+\tau_{4,2}x^2+\tau_{4,3}x+\tau_{4,4})y=0,
\end{align}
where $a_{6,0}^2+a_{5,0}^2\neq 0$ and at least one of the coefficients $a_{6,j},j=0,1,\dots,6$ is different from zero,
is
\begin{align}\label{eq11}
\tau_{4,0}=n(n-1)\cdot a_{6,0}+n\cdot a_{5,0}\qquad (n=0,1,2,\dots)
\end{align}
while the sufficient condition is given by the solution of the following system of linear equations 
\begin{equation}\label{eq12}
\left(
\begin{array}{llllllllllllll}
\alpha_0 & \beta_0 & \gamma_0& 0 & 0&0&0&0&\dots&0& 0&0&0&0\\
\delta_1 & \alpha_1 & \beta_1 &\gamma_1&0&0&0&0&\dots&0&0&0&0&0\\
\eta_2 & \delta_2 & \alpha_2&\beta_2&\gamma_2&0&0&0&\dots&0&0&0&0&0\\
\mu_3&\eta_3&\delta_3& \alpha_3& \beta_3&\gamma_3&0&0&\dots&0&0&0&0&0 \\
\zeta_4&\mu_4&\eta_4&\delta_4& \alpha_4& \beta_4&\gamma_4&0&\dots&0&0&0&0&0 \\
0&\zeta_5&\mu_5&\eta_5&\delta_5& \alpha_5& \beta_5&\gamma_5&\dots&0&0&0&0&0 \\
\vdots&\vdots&\vdots&\vdots&\vdots& \vdots&\vdots&\vdots&\ddots&\vdots&\vdots& \vdots&\vdots&\vdots \\
0&0&0&0&0&0&0&0&\dots&\eta_{n-2}&\delta_{n-2}&\alpha_{n-2}&\beta_{n-2}&\gamma_{n-2} \\
0&0&0&0&0&0&0&0&\dots&\mu_{n-1}&\eta_{n-1}&\delta_{n-1}&\alpha_{n-1}&\beta_{n-1} \\
0&0&0&0&0&0&0&0&\dots&\zeta_{n}&\mu_{n}&\eta_{n}&\delta_{n}&\alpha_{n}\\
\end{array}
\right)_{(n+1)\times(n+1)}\cdot 
\left(
\begin{array}{l}
c_0\\
c_1\\
c_2\\
c_3\\
c_4\\
c_5\\
\vdots\\
c_{n-2}\\
c_{n-1}\\
c_n\\
\end{array}
\right)_{(n+1)\times 1}
=0
\end{equation}
where
\begin{equation}\label{eq13}
\left\{ \begin{array}{lr}
\alpha_n=n(n-1)a_{6,4}+na_{5,4}-\tau_{4,4},&\qquad n=0,1,2,\dots\\ 
\beta_n=n(n+1)a_{6,5}+(n+1)a_{5,5},&\qquad n=0,1,2,\dots\\ 
\gamma_n=(n+2)(n+1)a_{6,6},&\qquad n=0,1,2,\dots\\ 
\delta_n=(n-1)(n-2)a_{6,3}+(n-1)a_{5,3}-\tau_{4,3},&\qquad n=1,2,\dots\\ 
\eta_n=(n-2)(n-3)a_{6,2}+(n-2)a_{5,2}-\tau_{4,2},& n=2,3,\dots \\ 
\mu_n=(n-3)(n-4)a_{6,1}+(n-3)a_{5,1}-\tau_{4,1},&\qquad n=3,4,\dots\\
\zeta_n=(n-4)(n-5)a_{6,0}+(n-4)a_{5,0}-\tau_{4,0},&\qquad n=4,5,\dots\\
       \end{array} \right.
\end{equation}
\end{theorem}
\noindent In particular, for a zero-degree polynomial solution of the differential equation \eqref{eq10} the sufficient and necessary conditions are
\begin{equation}\label{eq14}
\tau_{4,0}=\tau_{4,1}=\tau_{4,2}=\tau_{4,3}=\tau_{4,4}=0,
\end{equation}
while for a first-degree polynomial solution of equation \eqref{eq10}, we must have $\tau_{4,0}=a_{5,0}$ in addition to  vanishing the four $2\times 2$-determinants
\begin{align}\label{eq15}
&\left|
\begin{array}{ll}
-\tau_{4,4} & a_{5,5} \\
-\tau_{4,3} & a_{5,4}-\tau_{4,4}\\
\end{array}
\right|=0, ~\left|
\begin{array}{ll}
-\tau_{4,4} & a_{5,5} \\
-\tau_{4,2} & a_{5,3}-\tau_{4,3}\\
\end{array}
\right|=0,~\left|
\begin{array}{ll}
-\tau_{4,4} & a_{5,5} \\
-\tau_{4,1} & a_{5,2}-\tau_{4,2}\\
\end{array}
\right|=0,~
 \left|
\begin{array}{ll}
-\tau_{4,4} & a_{5,5} \\
-\tau_{4,0} & a_{5,1}-\tau_{4,1}\\
\end{array}
\right|=0. 
\end{align}
For a second-degree polynomial solution, we must have $\tau_{4,0}=2\cdot a_{6,0}+2\cdot a_{5,0}$
 in addition to  the vanishing of the four $3\times 3$-determinants
\begin{align}\label{eq16} 
&
\left|
\begin{array}{lllll}
-\tau_{4,4} & & a_{5,5} & & 2a_{6,6}\\
-\tau_{4,3} & & a_{5,4}-\tau_{4,4}& & 2a_{6,5}+2a_{5,5}\\
-\tau_{4,2} & & a_{5,3}-\tau_{4,3}&  &2a_{6,4}+2a_{5,4}-\tau_{4,4}
\end{array}
\right|=0,~\left|
\begin{array}{lllll}
-\tau_{4,4} & & a_{5,5} & & 2a_{6,6}\\
-\tau_{4,3} & & a_{5,4}-\tau_{4,4}& & 2a_{6,5}+2a_{5,5}\\
-\tau_{4,1} & & a_{5,2}-\tau_{4,2}& & 2a_{6,3}+2a_{5,3}-\tau_{4,3}
\end{array}
\right|=0,~\notag\\ \notag\\
&  \left|
\begin{array}{lllll}
-\tau_{4,4} & & a_{5,5} & & 2a_{6,6}\\
-\tau_{4,3} & & a_{5,4}-\tau_{4,4}& & 2a_{6,5}+2a_{5,5}\\
-\tau_{4,0} & & a_{5,1}-\tau_{4,1}& & 2a_{6,2}+2a_{5,2}-\tau_{4,2}
\end{array}
\right|=0,~
 \left|
\begin{array}{lllll}
-\tau_{4,4} & & a_{5,5} & &2a_{6,6}\\
-\tau_{4,3} & & a_{5,4}-\tau_{4,4}& &2a_{6,5}+2a_{5,5}\\
0 & & a_{5,0}-\tau_{4,0}& & 2a_{6,1}+2a_{5,1}-\tau_{4,1}
\end{array}
\right|=0.
\end{align}
Similar conditions follows for higher order of polynomial solutions. The first few polynomial solutions are given explicitly by
\begin{align}\label{eq17}
y_0(x)&=1,\qquad y_1(x)=1+\frac{\tau_{4,4}}{a_{5,5}}\cdot x,\notag\\
y_2(x)&=1+\frac{-a_{6,6}\tau_{4,3} + (a_{5,5} + a_{6,5}) \tau_{4,4}}{a_{5,5} (a_{5,5} + a_{6,5}) + a_{6,6} (-a_{5,4} + \tau_{4,4})}\cdot x+\frac12\cdot \frac{a_{5,5}\tau_{4,3} -a_{5,4}\tau_{4,4} + \tau_{4,4}^2}{(a_{5,5} (a_{5,5} + a_{6,5}) + a_{6,6} (-a_{5,4} + \tau_{4,4}))}\cdot x^2.
\end{align}
\vskip0.1true in
\noindent This theorem can be proved either by means of the  Frobenius series solution of ordinary Differential Equations or using the asymptotic iteration method \cite{saad1,saad8,saad7}.

%%%%%%%%%%%%%%%%%%%%%%%%%%%%%%%%%%%%%%%%%%%%%%%%%%%%%%%%%%%%%%%%%%%%%%%%%
\section{The one-dimensional decatic polynomial potentials}
%%%%%%%%%%%%%%%%%%%%%%%%%%%%%%%%%%%%%%%%%%%%%%%%%%%%%%%%%%%%%%%%%%%%%%%%%

\noindent For the existence of polynomial solutions of the differential equation \eqref{eq9}, it is clear from the criterion \eqref{eq11} that the parameters $a,b,c,$ and $d$ must be related through the equation
\begin{equation}\label{criterion}
8(5+2n)a^{5\over 2}+8a^2d-4abc+b^3=0,\qquad n=0,1,2,\dots
\end{equation}
There is an important consequence of this criterion that \emph{none} of the following decatic polynomials
\begin{align}\label{conseq}
V(x)=a\cdot x^{10},\quad V(x)=a\cdot x^{10}+e\cdot x^{2},\quad V(x)=a\cdot x^{10}+c\cdot x^{6},\quad V(x)=a\cdot x^{10}+c\cdot x^{6}+e\cdot x^{2},
\end{align}
have analytic solutions of the form \eqref{eq8} with $\chi(x)$ is a polynomial-type expression. Indeed, we can claim out of this criterion that none of these potentials \eqref{conseq} is a quasi-exactly solvable model \cite{ale} and we have to rely on perturbation or approximation methods to obtain the necessary information about the discrete spectrum of these potentials as is usually used in the literature  (e.g. \cite{bar,mee,cha,fra2,gom,kat,ko,er,moj,iva}). In the following, we shall focus on studying the decatic potentials \eqref{eq3}  using Theorem \ref{conds} to obtain the necessary conditions on the potential parameters for exact solutions followed by the application of the asymptotic iteration method to analyse the discrete spectrum for arbitrary values of the potential parameters.  

%%%%%%%%%%%%%%%%%%%%%%%%%%%%%%%%%%%%%%%%%%%%%%%%%%%%%%%%%%%%%%%%%%%%%%%%%
\subsection{Exact solutions}
%%%%%%%%%%%%%%%%%%%%%%%%%%%%%%%%%%%%%%%%%%%%%%%%%%%%%%%%%%%%%%%%%%%%%%%%%

\noindent For the necessary condition for even-degree polynomial solutions of Schr\"odinger equation \eqref{eq3} we have, by means of equation \eqref{eq11}  that
\begin{equation}\label{eq18}
8(4n+5) a^{5\over 2}+8a^2d-4abc+b^3=0,\qquad n=0,1,2,\dots
\end{equation}
while for the sufficient condition we deduce, by means of equations \eqref{eq12}-\eqref{eq13}, that
\begin{align}\label{eq19}
	\chi_{2n+2}(x)=\sum_{k=0}^{n}\frac{(-1)^k\cdot P_{2k}(E)}{(2k)!\cdot (2\sqrt{a})^{3k}}\cdot x^{2k}, \qquad (n=0,1,2,\dots).
\end{align}
where the polynomials $P_{2k}(E)$ satisfies the four-term recurrence relation
\begin{align}\label{eq20}
	P_{2n+2}(E)		&\equiv		\Big(8a^{3/2}E+(4n+1)\left(b^2-4ac\right)\Big)\cdot P_{2n}(E)\notag\\
					&		+2n\cdot (2n-1)\cdot (32(4n-1)a^{5/2}b-b^4+8ab^2c-16 a^2 c^2+64a^3e)\cdot P_{2n-2}(E)\notag\\
&+16384\cdot a^5\cdot n\cdot (n-1)\cdot (2n-1)\cdot (2n-3)\cdot P_{2n-4}(E), 
\end{align}
initiated with $P_0(E)=1, P_{-n}(E)=0,~n\geq 1$. The exact (real) eigenenergies $E$ are the roots of equation ${P}_{2n+2}(E)= 0$ subject to the following additional condition
\begin{equation}\label{eq21}
	\Big(32(4n+3)a^{5/2}b-b^4+8ab^2c-16a^2c^2+64a^3e\Big)\cdot P_{2n}(E)+4096\cdot a^5 \cdot n\cdot (2n-1)\cdot P_{2n-2}(E)=0,\qquad (n=0,1,2,\dots).
  \end{equation}
As an example, for the ground state eigenvalue $n=0$, we have for arbitrary  values of the potential parameters $a,~b,$ and $c$ that
\begin{equation}\label{eq22}
E_0 = \frac{1}{8}\cdot\frac{4ac-b^2}{a^{{3}/{2}}},~ d = -\frac{1}{8}\cdot\frac{40a^{5/2}+b^3-4abc}{a^2},~ e = -\frac{1}{64}\cdot \frac{96a^{5/2}b-b^4+8ab^2c-16a^2c^2}{a^3}
\end{equation}

\noindent On the other hand, the necessary condition for odd-degree polynomial solutions of Schr\"odinger equation \eqref{eq3}, by means of equation \eqref{eq11}, is
\begin{equation}\label{eq23}
8(4n+7) a^{5\over 2}+8a^2d-4abc+b^3=0,\qquad (n=0,1,2,\dots)
\end{equation}
under which the exact wavefunctions are given by 
\begin{equation}\label{eq24}
\chi_{2n+3}(x)=\sum_{k=0}^{n}\frac{(-1)^k\cdot P_{2k+1}(E)}{(2k+1)!\cdot (2\sqrt{a})^{3k}}x^{2k+1}, \qquad n=0,1,2,\dots
\end{equation}
where the polynomials  $P_{2k+1}(E)$ are given explicitly using the recurrence relation 
\begin{align}\label{eq25} 
	{P}_{2n+3}(E)		&\equiv		\Big(8a^{3/2}E+(4n+3)\left(b^2-4ac\right)\Big)\cdot {P}_{2n+1}(E)\notag \\
					&		+2n\cdot (2n+1)\cdot\Big(32(4n+1)a^{5/2}b-b^4+8ab^2c-16a^2c^2+64a^3e\Big)\cdot {P}_{2n-1}(E)\notag\\
					&		+16384\cdot a^5\cdot n\cdot (n-1)\cdot (2n-1)\cdot (2n+1)\cdot {P}_{2n-3}(E),\qquad (n=0,1,2,\dots)
\end{align}
initiated with $P_1(E)=1$ and $P_{-n}=0$ for $n\geq 1$.  The exact eigenenergies $E$ are the real roots of ${P}_{2n+3}(E)= 0$, subject to the additional condition 
\begin{equation}\label{eq26}
\Big(32 (4n+5) a^{5/2} b-b^4+8ab^2c-16a^2c^2+64a^3e\Big)\cdot {P}_{2n+1}(E)+4096\cdot a^5 \cdot n\cdot (2n+1)\cdot {P}_{2n-1}(E)=0,\qquad (n=0,1,2,\dots).
  \end{equation}
As an example, for the first excited  state $n=1$, we have for the potential parameters $a,~b,$ and $c$ that
\begin{equation}\label{eq27}
E_1 = \frac{3}{8}\cdot\frac{4ac-b^2}{a^{{3}/{2}}},~ d = -\frac{1}{8}\cdot\frac{56a^{5/2}+b^3-4abc}{a^2},~ e = -\frac{1}{64}\cdot \frac{160a^{5/2}b-b^4+8ab^2c-16a^2c^2}{a^3}
\end{equation}

\noindent The first few polynomial solutions, $n=0,1,\dots,4$, are reported explicitly in the appendix. In Table \ref{T1},  we report the ground-state eigenvalue for the decatic potentials \eqref{eq3} for specific values of the potential parameters $a,~b,~c,~d$ and $e$. Further, in Table \ref{T2}, we report the first-excited state for particular values of the potential parameters.  These results confirm and generalize the early finding by Chaudhuri and Mondal \cite{cha}, using the improved Hill determinant method, for the particular potentials:
\begin{align}
V(x) &= \dfrac{105}{64}x^2-\dfrac{43}{8}x^4+x^6-x^8+x^{10},\qquad \left(E_0=\dfrac{3}{8},~\psi_0(x)=\exp\left({-\dfrac{3}{16}\cdot x^2+\dfrac{1}{8}\cdot x^4-\dfrac{1}{6}\cdot x^6}\right)\right),\label{eq28}\\
V(x) &= \dfrac{169}{64}x^2-\dfrac{59}{8}x^4+x^6-x^8+x^{10},\qquad \left(E_1=\dfrac{9}{8},~\psi_1(x)=x\cdot \exp\left({-\dfrac{3}{16}\cdot x^2+\dfrac{1}{8}\cdot x^4-\dfrac{1}{6}\cdot x^6}\right)\right).\label{eq29}
\end{align}

\begin{figure}[htb!]
\centering%
\includegraphics[width=80mm]{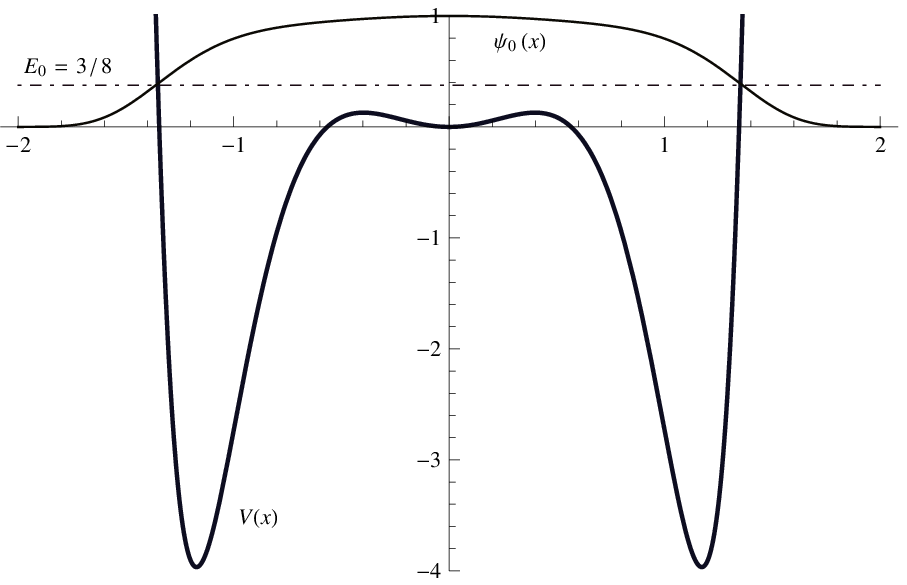}\hskip0.5true in\includegraphics[width=80mm]{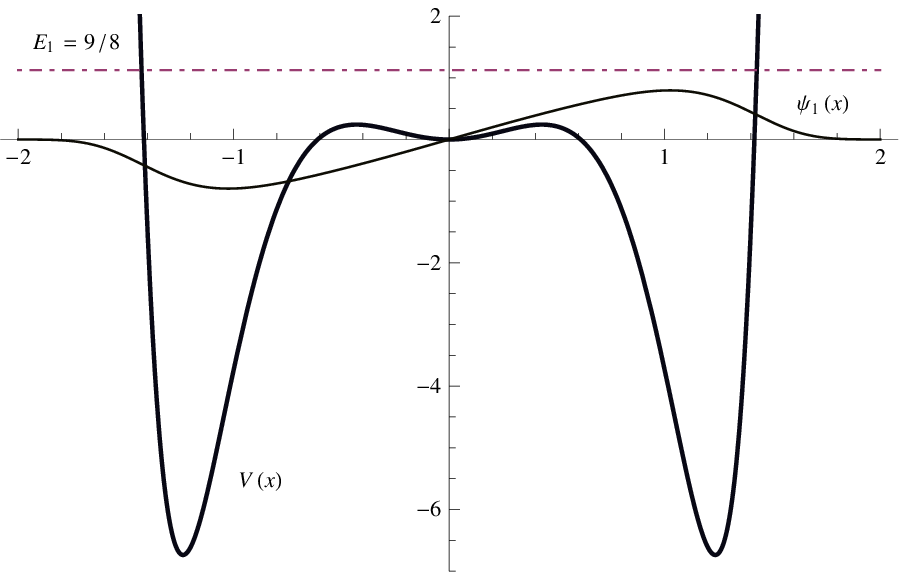}
\caption{The ground-state $n=0$ and first-excited state  $n=1$ (un-normalized) wave functions for the decatic potential energy $V(x)=x^{10}+x^8+x^6+d x^4+ e x^2$ along with the exact eigenvalues as are given by \eqref{eq28} and \eqref{eq29}, respectively.}
\label{Fig1}
\end{figure}

\noindent The ground-state $n=0$ and first-excited state $n=1$ along with the un-normalized wave functions for the potentials \eqref{eq28} and \eqref{eq29} are displayed in Figure \ref{Fig1}. 

\begin {table}[!h]
\begin{center}
\begin{tabular}{|c|c|c|c|c|c|c|}
\hline
 $E_0$ & $a$ & $b$ & $c$  & $d$ & $e$\\
\hline
${3}\sqrt{\mu}/8$~&~$\mu$ & $\pm \mu$ & $\mu$ & $\frac{1}{8}\cdot(\pm 3\mu-40\sqrt{\mu})$~ &~ $\frac{3}{64}\cdot(3\mu\mp32\sqrt{\mu})$~\\
\hline
$-5\sqrt{\mu}/8$~&~$\mu$ & $\pm \mu$ & $-\mu$ & $\mp\frac{5}{8}\cdot(\mu\pm 8\sqrt{\mu})$~ &~ $\frac{1}{64}\cdot(25\mu\mp96\sqrt{\mu})$~\\
\hline
${(4k-1)}\sqrt{k\mu}/(8k^2)$~&$k\cdot\mu$ & $\mu$ & $\mu$ & $ -\frac{1}{8k^2}\cdot {(40k^2\sqrt{k\mu}+(1-4k)\mu)}$~ &~ $ \frac{1}{64k^3}\cdot ({(1-4k)^2\mu-96k^2\sqrt{k\mu}})$~\\
\hline
$-(k^2-4)\sqrt{\mu}/8$~&$\mu$ & $k\cdot \mu$ & $\mu$ & $ -\frac{1}{8}\cdot k\cdot(k^2-4)\cdot \mu-5\cdot \sqrt{\mu}$~ &~ $ \frac{1}{64}\cdot (k^2-4)^2\cdot \mu-\frac32\cdot k\cdot \sqrt{\mu}~$~\\
\hline
$(4k-1)\sqrt{\mu}/8$~&$\mu$ & $\mu$ & $k\cdot \mu$ & $ \frac18\cdot (4k-1)\cdot \mu-5\cdot \sqrt{\mu}$~ &~ $ \frac{1}{64}\cdot (1-4k)^2\mu-\frac32\cdot \sqrt{\mu}~$~\\
\hline
$\sqrt{k\cdot \mu}+5\cdot k$~&$k\cdot \mu$ & $\mu$ & $\frac{\mu}{4k} +2k\cdot (\mu+5\sqrt{\mu})$ & $\mu$~ &~ $25k^2-\dfrac{3}{2}\sqrt{\dfrac{\mu}{k}}+k(\mu+10\sqrt{k\mu})$~\\
\hline
$(\sqrt{\mu}+5)/k$~&$\mu$ & $k\cdot \mu$ & $\frac{\sqrt{\mu}}{4k}\cdot ({40 +\sqrt{\mu}\cdot (8+k^3)})$ & $\mu$~ &~ $\dfrac{1}{2k^2}(2\mu+(20-3k^3)\sqrt{\mu}+50)$~\\
\hline
$5+{k\cdot \sqrt{\mu}}$~&$\mu$ & $\mu$ & $\frac{\sqrt{\mu}}{4}\cdot ({40+\sqrt{\mu}\cdot (1+8\cdot k)})$ & $k\cdot \mu$~ &~ $k^2\cdot \mu+(10\cdot k-\frac32)\cdot\sqrt{\mu}+25$~\\
\hline
\end {tabular}
\caption{The ground-state energy $E_{0}$  for a decatic-power potential $V(x)=a\cdot x^{10}+b\cdot x^{8}+c\cdot x^6+d\cdot x^4+e\cdot x^2$ for different values of the parameters. Here, $\mu$ and $k$ are arbitrary positive numbers.}\label{T1}
\end{center}
\end {table}

\begin{figure}[htb!]
\centering%
\includegraphics[width=80mm]{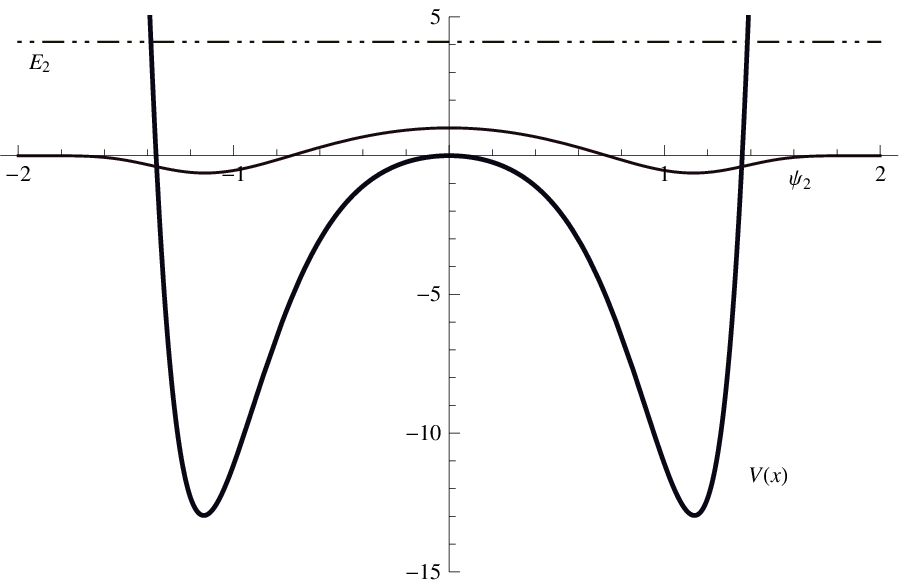}\hskip0.5true in\includegraphics[width=80mm]{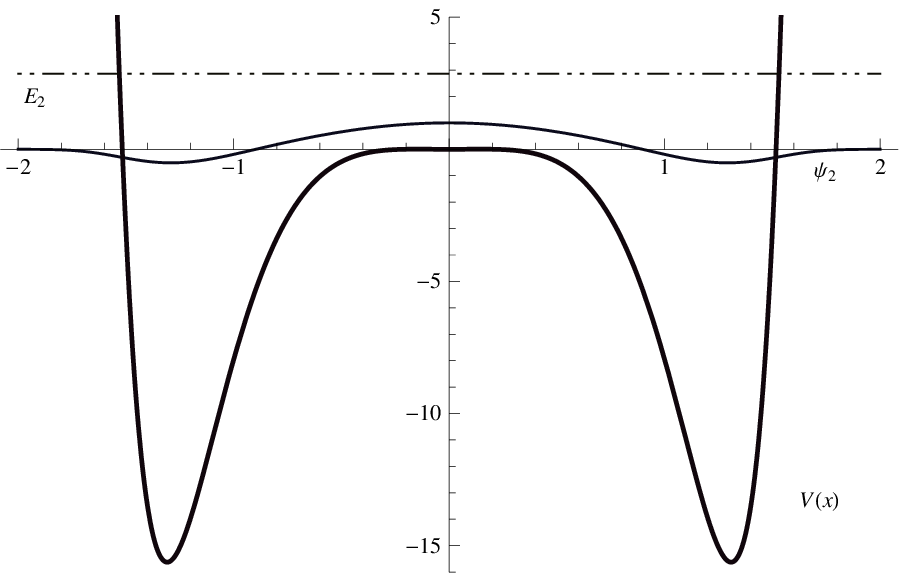}
\caption{The decatic potentials \eqref{eq30} and \eqref{eq33}, respectively, and their exact solutions.}
\label{Fig2}
\end{figure}

\begin {table}[!h]
\begin{center}
\begin{tabular}{|c|c|c|c|c|c|c|}
\hline
 $E_1$ & $a$ & $b$ & $c$  & $d$ & $e$\\
\hline
${9}\sqrt{\mu}/8$~&~$\mu$ & $\pm\mu$ & $\mu$ & $\frac{1}{8}\cdot(\pm 3\mu-56\sqrt{\mu})$~ &~ $\frac{1}{64}\cdot(9\mu\mp 160\sqrt{\mu})$~\\
\hline
$-{15}\sqrt{\mu}/8$~&~$\mu$ & $\pm\mu$ & $-\mu$ & $\frac{1}{8}\cdot(\mp 5\mu-56\sqrt{\mu})$~ &~ $\frac{5}{64}\cdot(5\mu\mp 32\sqrt{\mu})$~\\
\hline
${3}{(4k-1)}\sqrt{k\cdot\mu}/(8k^2)$~&$k\cdot\mu$ & $\mu$ & $\mu$ & $ -\frac{1}{8k^2}\cdot ({56k^2\sqrt{k\mu}+\mu-4k\mu})$~ &~ $ \frac{1}{64k^3}({-160k^2\sqrt{k\mu}+(1-4k)^2\mu})$~\\
\hline
$-\frac{3}{8}(k^2-4)\sqrt{\mu}$~&$\mu$ & $k\cdot \mu$ & $\mu$ & $ -\frac{k}{8}(k^2-4)\mu-7\cdot \sqrt{\mu}$~ &~ 
$\frac{1}{64}(1-4k)^2\mu-\frac52\cdot k\cdot \sqrt{\mu}~$~\\
\hline
$\frac38(4k-1)\sqrt{\mu}$~&$\mu$ & $\mu$ & $k\cdot \mu$ & $ -\frac18\cdot \mu+\frac12\cdot k\cdot \mu-7\cdot \sqrt{\mu}$~ &~ $ \frac{1}{64}\cdot \mu-\frac18\cdot k\cdot \mu+\frac14\cdot k^2\cdot \mu-\frac52\cdot \sqrt{\mu}~$~\\
\hline
$3\sqrt{k\cdot \mu}+21\cdot k$~&$k\cdot \mu$ & $\mu$ & $\frac14\cdot \frac{56\cdot (k\cdot \mu)^{5/2}+\mu^3\cdot (1+8\cdot k^2)}{k\cdot \mu^2}$ & $\mu$~ &~ $\frac12\cdot \frac{-5\cdot\sqrt{k\cdot\mu}+98k^3+28\cdot k^2\cdot \sqrt{k\cdot\mu}+2\cdot k^2\cdot \mu}{k}$~\\
\hline
${3(\sqrt{\mu}+7)}/{k}$~&$\mu$ & $k\cdot \mu$ & $\frac{\sqrt{\mu}}{4k}\cdot({56+\sqrt{\mu}\cdot (k^3+8)})$ & $\mu$~ &~ $-\frac1{2k^2}\cdot (98+(28-5k^3)\sqrt{\mu}+2\mu)$~\\
\hline
${3(7+k\cdot \sqrt{\mu})}$~&$\mu$ & $\mu$ & $\frac14\cdot (56\cdot \sqrt{\mu}+\mu\cdot (1+8\cdot k))$ & $k\cdot \mu$~ &~ $k^2\cdot \mu+14\cdot k\cdot\sqrt{\mu}-\frac52\cdot\sqrt{\mu}+49$~\\
\hline
\end {tabular}
\caption{The first excited state energy $E_{1}$  for a decatic-power potential. Here, $\mu$ and $k$ are arbitrary positive numbers.}\label{T2}
\end{center}
\end {table}
\noindent For higher states, we note, using the results reported in the appendix or using Theorem \ref{conds} directly, for the potential
\begin{align}\label{eq30}
V(x)&=x^{10}+x^8+x^6-\dfrac{69}{8}x^4\notag\\
&+\dfrac{24 \left(1971 - 24 \sqrt{6423}\right)^{1\over 3} - 
   8 (1971 - 24 \sqrt{6423})^{2\over 3} + 
   24 (1971 + 24 \sqrt{6423})^{1\over 3} - 
   8 (1971 + 24 \sqrt{6423})^{2\over 3}-1551}{576}x^2,
\end{align}
the exact eigenvalue and wave function are given, in the case $n=2$, by 
\begin{align}
E_2&=\dfrac{1}{24} \left(21 + 4 (1971 - 24 \sqrt{6423}))^{1\over 3} + 
   4 (1971 + 24\sqrt{6423}))^{1\over 3}\right)\label{eq31}\\
   \psi_2(x)&=\left(1-\dfrac{1}{12}\left({21 + 4 (1971 - 24 \sqrt{6423}))^{1\over 3} + 
   4 (1971 + 24\sqrt{6423}))^{1\over 3}}\right)\cdot x^2\right)\cdot \exp\left(-\dfrac{1}{6}\cdot x^6-\dfrac{1}{8}\cdot x^4-\dfrac{3}{16}\cdot x^2\right).\label{eq32}
\end{align}
Further, for the potential 
\begin{align}\label{eq33}
V(x)&=x^{10}-x^8+x^6-\dfrac{75}{8}\cdot x^4\notag\\
&+\dfrac{64{\root 3\of 3}(43+\sqrt{4215})-64{\root 6\of 3}(23\sqrt{3}+\sqrt{1405})(513+8\sqrt{4215})^{1\over 3}+571(513+8\sqrt{4215})^{2\over 3}}{192(513+8\sqrt{4215})^{2\over 3}}\cdot x^2,
\end{align}
we have the exact solutions 
\begin{align}
E_2&=\dfrac{1}{8} \left(7 + \dfrac{4}{3^{2\over 3}}\cdot(513+8 \sqrt{4215})^{1\over 3} - 
   \dfrac{52}{3 (513 + 8\sqrt{4215}))^{1\over 3}}\right)\label{eq34},\\
   \psi_2(x)&=\left(1-\dfrac{1}{16}\left({4 + \dfrac{4}{3^{2\over 3}}\cdot(513+8 \sqrt{4215})^{1\over 3} - 
   \dfrac{52}{3 (513 + 8\sqrt{4215}))^{1\over 3}}}\right)\cdot x^2\right)\cdot \exp\left(-\dfrac{1}{6}\cdot x^6-\dfrac{1}{8}\cdot x^4-\dfrac{3}{16}\cdot x^2\right).\label{eq35}
\end{align}
The plots of these potentials along with their exact solutions for $n=2$ are displayed in figure \ref{Fig2}.

%%%%%%%%%%%%%%%%%%%%%%%%%%%%%%%%%%%%%%%%%%%%%%%%%%%%%%%%%%%%%%%%%%%%%%%%%%%%%%%%%%%%%%%%%
\subsection{Approximate solutions}
\begin{figure}[htb!]
\centering%
\includegraphics[width=80mm]{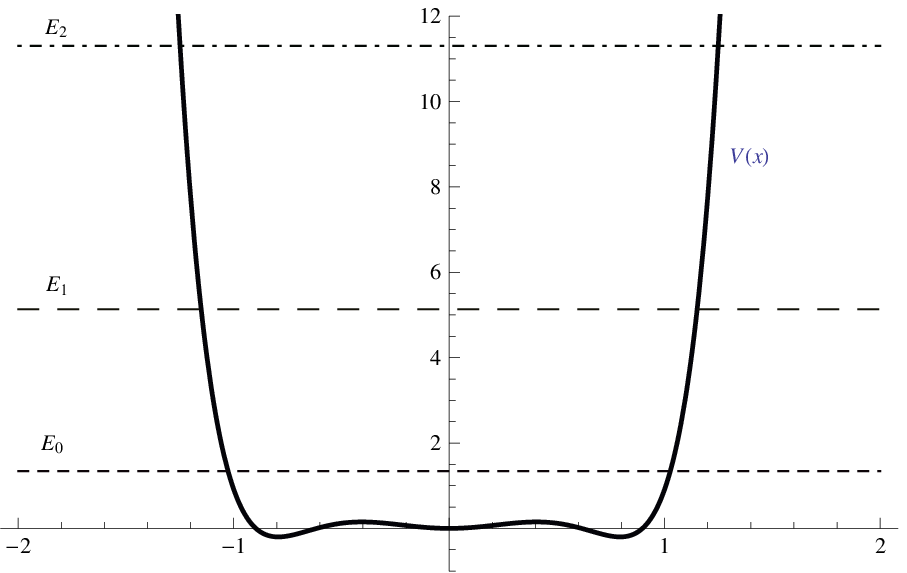}\hskip0.5true in\includegraphics[width=80mm]{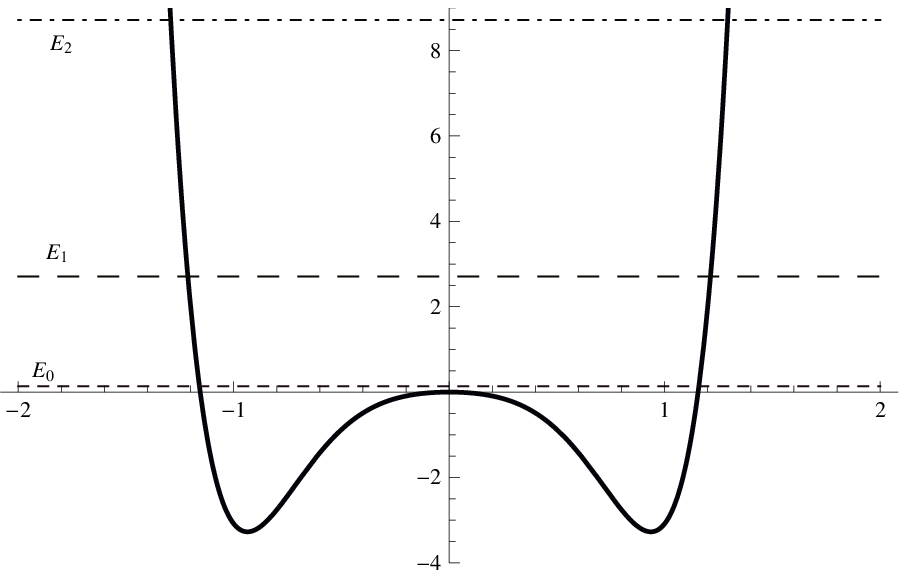}
\caption{The graph of the potentials $V(x)=0.04x^{10}+0.877x^8+5.5x^6-7.5x^4+2x^2$ and $ V(x)=0.04x^{10}+0.877x^8+5.5x^6-7.5x^4-2x^2$, respectively.}
\label{Fig3}
\end{figure}

%%%%%%%%%%%%%%%%%%%%%%%%%%%%%%%%%%%%%%%%%%%%%%%%%%%%%%%%%%%%%%%%%%%%%%%%%%%%%%%%%%%%%%%%%
\noindent The asymptotic iteration method was introduced \cite{saad1} to investigate the
solutions of differential equations of the form
\begin{equation}\label{eq36}
y''=\lambda_0(x) y'+s_0(x) y,\quad\quad ({}^\prime={d\over dx})
\end{equation}
where $\lambda_0\equiv\lambda_0(x)$ and $s_0\equiv s_0(x)$ are $C^{\infty}-$differentiable functions.
A key feature of this method is to note the invariant structure of the right-hand
side of (\ref{eq36}) under further differentiation. Indeed, if we differentiate (\ref{eq36}) with respect to $x$, we obtain
\begin{equation}\label{eq37}
y^{\prime\prime\prime}=\lambda_1\cdot y^\prime+s_1\cdot y
\end{equation}
where $\lambda_1= \lambda_0^\prime+s_0+\lambda_0^2$ and $s_1=s_0^\prime+s_0\lambda_0.$
If we find the second derivative of equation (\ref{eq36}), we obtain
\begin{equation}\label{eq38}
y^{(4)}=\lambda_2\cdot y^\prime+s_2\cdot  y
\end{equation}
where $\lambda_2= \lambda_1^\prime+s_1+\lambda_0\lambda_1$ and $s_2=s_1^\prime+s_0\lambda_1.$
Thus, for $(n+1)^{th}$ and $(n+2)^{th}$ derivative of (\ref{eq36}), $n=1,2,\dots$, we have
\begin{equation}\label{eq39}
y^{(n+1)}=\lambda_{n-1}\cdot y^\prime+s_{n-1}\cdot y
\end{equation}
and
\begin{equation}\label{eq40}
y^{(n+2)}=\lambda_{n}\cdot y^\prime+s_{n}\cdot y
\end{equation}
respectively, where
\begin{equation}\label{eq41}
\lambda_{n}= \lambda_{n-1}^\prime+s_{n-1}+\lambda_0\lambda_{n-1}\hbox{ ~~and~~ } s_{n}=s_{n-1}^\prime+s_0\lambda_{n-1}.
\end{equation}
From (\ref{eq39}) and (\ref{eq40}) we have
\begin{equation}\label{eq42}
\lambda_n y^{(n+1)}- \lambda_{n-1}y^{(n+2)} = \delta_ny {\rm ~~~where~~~}\delta_n=\lambda_n s_{n-1}-\lambda_{n-1}s_n.
\end{equation}
Clearly, from (\ref{eq42}) if $y$, the solution of (\ref{eq36}), is a polynomial of degree $n$, then $\delta_n\equiv 0$. Further, if $\delta_n=0$, then $\delta_{n'}=0$ for all $n'\geq n$. In the original  paper on the Asymptotic Iteration Method (AIM)  \cite{saad1}, the following  theorem was proved.
\vskip0.1in
\begin{theorem} Given $\lambda_0$ and $s_0$ in $C^{\infty}(a,b),$
the differential equation (\ref{eq36}) has the general solution
\begin{equation}\label{eq43}
y(r)= \exp\left(-\int\limits^{r}\alpha(t) dt\right)
\left[C_2 +C_1\int\limits^{r}\exp\left(\int\limits^{t}(\lambda_0(\tau) + 2\alpha(\tau)) d\tau \right)dt\right]
\end{equation}
if for some $n>0$
\begin{equation}\label{eq44}
{s_{n}\over \lambda_{n}}={s_{n-1}\over \lambda_{n-1}} \equiv \alpha\qquad\mbox{or equivalently}\quad \delta_n\equiv s_n\cdot \lambda_{n-1}-s_{n-1}\cdot \lambda_n=0.
\end{equation}
\end{theorem}
\vskip0.1true in
\noindent The method can apply directly to the differential equation \eqref{eq9}. We have considered here, as an example of the effectiveness of the method, the following two decatic potentials (see figure \ref{Fig3})
\begin{align}\label{eq45}
V(x)&=0.04x^{10}+0.877x^8+5.5x^6-7.5x^4+2x^2,
\end{align}
and
\begin{align}\label{eq46}
V(x)&=0.04x^{10}+0.877x^8+5.5x^6-7.5x^4-2x^2.
\end{align}
 
\noindent For these specific values of the potential parameters \eqref{eq9}, the AIM sequences $\lambda_n$ and $s_n$, $n=1,2,\dots$, can be initiated with 
\begin{align}\label{eq47}
\left\{ \begin{array}{l}
\lambda_0(x)=2\sqrt{a}x^5+\frac{b}{\sqrt{a}}x^3-\frac{\left(b^2-4ac\right)}{4a^{3/2}}x,\\ \\
s_0(x)=\dfrac{4ac-b^2}{8a^{3/2}}-E+\dfrac{64a^3e+96a^{5/2}b-b^4+8ab^{2}c-16a^2c^2}{64a^3}x^2+
				\dfrac{8a^2d+40a^{5/2}+b^3-4abc}{8a^2}x^4.
       \end{array} \right.
\end{align}
and by means of the iteration sequences \eqref{eq35}, we compute $\lambda_n(x)$ and $s_n(x),$ for $n=1,2,3,\dots$. The 
eigenvalues are then the roots of the termination condition \eqref{eq44}, namely $\delta_n(x_0;E) = 0$. With several symbolic mathematical programs available (Maple; Mathematica, etc.), the computation of the roots of this equation, and thus the eigenvalues, by means of the
iteration method, provided it is set up correctly, is a straightforward calculation,
even for the higher iteration steps as shown in Table \ref{T5}. In principle, the computation of the roots of $\delta_n(x_0;E) = 0$ should be independent of the choice of $x=x_0$,
nevertheless, the right choice of $x_0$, usually accelerates the rate of convergence to
accurate eigenvalues within a reasonable number of iterations  \cite{bro}. For our computation, we have fixed the initial value of $x$ as $x_0=0$.

\begin {table}[!ht]
\begin{center}
\begin{tabular}{||c|c||c|c||}
\hline
$n$ & $E_{n}(a=0.04,b=0.877,c=5.5,d=-7.5,e=2)$&$n$&$E_{n}(a=0.04,b=0.877,c=5.5,d=-7.5,e=-2)$\\ 
\hline
0     &~  $~1.342~322~779~564~646~216_{189}$~&~$0$~&~$0.135~429~448~914~739~200_{187}$ \\
\hline
1    &~  $~5.136~482~243~202~476~766_{196}$~&~$1$~&~$2.708~255~201~038~304~023_{194}$ \\
\hline
2    &~  $11.304~205~821~188~349~432_{203}$~&~$2$~&~$8.719~471~783~294~745~896_{199}$  \\
\hline
3&~$19.577~677~092~617~956~963_{208}$~&~$3$~& $ 16.684~057~789~758~348~655_{204}$\\
\hline
4&~$29.438~055~052~333~889~749_{215}$~&~$4$~& $26.234~813~143~437~435~201_{213}$ \\
\hline
5&~$40.723~083~272~298~272~621_{222}$~&~$5$~& $37.249~467~272~037~347~758_{234} $  \\
\hline
\hline
$n$ & $E_{n}(a=0.01,b=0.1,c=1.0,d=3.250,e=12.5625) $&$n$&$E_{n}(a=0.01,b=0.1,c=1.0,d=3.050,e=11.5625)$\\ 
\hline
0     &~  $~3.750~000~000~000~000~000_{6}^{Exact}$~&~$0$~& $3.611~850~704~712~528~538_{83}$ \\
\hline
1    &~  $~11.653~048~680~350~115~469_{104}$~&~$1$~&~ $11.250~000~000~000~000~000_{6}^{Exact}$ \\
\hline
2    &~  $20.358~948~672~177~131~878_{113}$~&~$2$~& $19.713~587~721~031~462~373_{109}$  \\
\hline
3&~$29.850~701~419~298~223~478_{130}$~&~$3$~& $28.983~552~585~573~622~185_{122}$\\
\hline
4&~$40.098~623~827~649~000~672_{139}$~&~$4$~& $39.026~595~688~356~661~175_{143}$ \\
\hline
5&~$51.071~414~767~832~192~856_{152}$~&~$5$~& $49.808~260~402~594~522~700_{146} $  \\
\hline
\end {tabular}
\caption {The first six energy eigenvalues (in units of $2m=\hbar=1$) for the one-dimensional decatic potentials $V(x)=ax^{10}+bx^8+cx^6+dx^4+ex^2$ for various values of parameters, calculated using AIM with $x_0=0$. The energies are accurate for the number of digits reported. The subscript refer to the number of iteration used by AIM.}\label{T5}
\end{center}
\end {table}

%%%%%%%%%%%%%%%%%%%%%%%%%%%%%%%%%%%%%%%%%%%%%%%%%%%%%%%%
\section{Conclusion}

\noindent We have discussed exact and approximate solutions of Schr\"odinger's 
equations with even-power polynomial potentials. Our analysis of exact solutions is based on introducing necessary and sufficient conditions of the polynomial solution for a class of differential equations with polynomial coefficients.  This analysis can serve as available tools to study more complicated physical systems implied by the differential equation. Although we confined our discussion in the present work to the decatic potential, our approach can be used to analyse different classes of oscillators with higher anharmonicities.  For the arbitrary values of the potential parameters, we have applied the asymptotic iteration method to solve Schr\"odinger's equation for which the method proves to be extremely
effective for finding the approximate solutions. We hope that the study presented here encourages further analysis of (both confined and unconfined) polynomial potentials.
%% ------------------------------------------------------
\section*{Acknowledgements}
%% ------------------------------------------------------
%\medskip
\noindent Partial financial support of this work under Grant No.GP249507 from the Natural Sciences and Engineering
Research Council of Canada is gratefully acknowledged.

%% ------------------------------------------------------
\section*{Appendix}
%% ------------------------------------------------------
The first few polynomial solutions of the decatic power potentials using the recurrence relations \eqref{eq20} amd \eqref{eq24} are:
\begin{itemize}
	\item For a zero-degree polynomial solution , $n=0$ and $\chi_2(x)=1$, 
	\begin{enumerate}
		\item $P_{2}(E):=8a^{3/2}E+b^2-4ac=0$,
		\item $96 a^{5/2} b-b^4+8 a b^2 c-16 a^2 c^2+64 a^3 e=0$,
		\item $40a^{5/2}+b^3-4 a b c+8 a^2 d=0.$		
	\end{enumerate}
  
	\item For a first-degree polynomial solution, $n=1$ and $\chi_3(x)=c_0+c_1x$, we have  $c_0=0$, and
	\begin{enumerate}
		\item $P_{3}(E):=8a^{3/2}E+3(b^2-4ac)=0$,		
		\item $160a^{5/2} b-b^4+8ab^2c-16 a^2c^2+64 a^3e=0$,
		\item $56a^{5/2}+b^3-4 a b c+8 a^2 d=0$.
	\end{enumerate}
Thus, if $c_1=1$, 
\begin{equation*}		
		\chi_3(x)=x.
\end{equation*}
	\item For a second-degree polynomial solution, $n=2$ and $\chi_4(x)=c_0+c_1x+c_2x^2$,	 we have $c_1=0$ and 
	\begin{enumerate}
		\item $P_{4}(E):=\big(8a^{3/2}E+5(b^2-4ac)\big)P_2(E)+2\big(96a^{5/2}b-b^4+8ab^2c-16a^2c^2+64a^3e\big)P_0(E)=0$,
		\item $\big(224 a^{5/2}b-b^4+8ab^2c-16a^2c^2+64a^3e\big)P_2(E)+4096a^5P_0(E)=0$,
		\item $72a^{5/2}+b^3-4 a b c+8 a^2 d=0$.
	\end{enumerate}
	Further,  with $c_1=0$, 	$c_2=-({8a^{3\over 2} E+b^2-4ac})c_0/({16 a^{3\over 2}})$,
	and for 		$c_0=1$
		we have
\begin{equation*}		
		\chi_4(x)=P_0(E)- \frac{P_2(E)}{16a^{3/2}}x^2.
\end{equation*}

	\item For a third-degree polynomial solution, $n=3$ and $\chi_5(x)=c_0+c_1x+c_2x^2+c_3x^3$, we have $c_0=c_2=0$ and further 
	\begin{enumerate}
		\item $P_{5}(E):=\big(8a^{3/2}E+5(b^2-4ac)\big)P_3(E)+6\big(160a^{5/2} b-b^4+8ab^2c-16 a^2c^2+64 a^3e\big)P_1(E)=0$
		\item $\big(288a^{5/2} b-b^4+8ab^2c-16 a^2c^2+64 a^3e\big)P_3(E)+12288a^5=0$
		\item $88a^{5/2}+b^3-4 a b c+8 a^2 d=0$
		\end{enumerate}
		In addition, for $c_1=1$,  $c_3=-{P_3(E)}/({48a^{3/2}})$, and the polynomial solution is
\begin{equation*}		
		\chi_5(x)=x\left(P_1(E)- \dfrac{P_3(E)}{48a^{3/2}}x^2\right).
\end{equation*}
	
	\item For a fourth-degree polynomial solution,  $n=4$ and $\chi_6(x)=c_0+c_1x+c_2x^2+c_3x^3+c_4x^4$, we have	$c_1=c_3=0$ and 
	\begin{enumerate}
		\item $P_{6}(E):=\big(8a^{3/2}E+9(b^2-4ac)\big)P_4(E)+12\big(224 a^{5/2}b-b^4+8ab^2c-16a^2c^2+64a^3e\big)P_2(E)+$\\
			\qquad$+1536a^{5/2}\big(40a^{5/2}+b^3-4abc+8a^2d\big)P_0(E)=0$
			\item $\big(352a^{5/2}b-b^4+8ab^2c-16a^2c^2+64a^3e\big)P_4(E)+24576a^5P_2(E)=0$
		\item $104a^{5/2}+b^3-4 a b c+8 a^2 d=0$
	\end{enumerate}
	In addition, with $c_0=1$, $c_2=-\dfrac{P_2(E)}{16a^{3/2}}$ and $c_4=\dfrac{P_4(E)}{1536a^3}$, and the polynomial solution is 
	\begin{equation*}		
		\chi_6(x)=P_0(E)- \dfrac{P_2(E)}{16a^{3/2}}x^2+ \dfrac{P_4(E)}{1536a^3}x^4.
\end{equation*}
\end{itemize}
%%%%%%%%%%%%%%%%%%%%%%%%%%%%%%%%%%%%%%%%%%%%%%%%%%%%%%%%%%%

\end{document}